# 4GREAT - a four-color receiver for high-resolution airborne terahertz spectroscopy

Carlos A. Durán, Rolf Güsten, Christophe Risacher, Andrej Görlitz, Bernd Klein, Nicolas Reyes, Oliver Ricken, Hans-Joachim Wunsch, Urs U. Graf, Karl Jacobs, Cornelia E. Honingh, Jürgen Stutzki, Gert de Lange, Yan Delorme, Jean-Michel Krieg, Dariusz C. Lis

*Abstract* —**4GREAT is an extension of the German Receiver for Astronomy at Terahertz frequencies (GREAT) operated aboard the Stratospheric Observatory for Infrared Astronomy (SOFIA). The spectrometer comprises four different detector bands and their associated subsystems for simultaneous and fully independent science operation. All detector beams are co-aligned on the sky. The frequency bands of 4GREAT cover 491–635, 890–1090, 1240–1525 and 2490–2590 GHz, respectively. This paper presents the design and characterization of the instrument, and its in-flight performance. 4GREAT saw first light in June 2018, and has been offered to the interested SOFIA communities starting with observing cycle 6.**

## I. Introduction

SINCE the completion of the Herschel satellite mission [1], the spectrometer GREAT[1] [2] on board the SOFIA airborne observatory [3] is the only instrument that routinely allows performing high-resolution spectroscopy of astronomical signals at far-infrared wavelengths. Except in very limited windows (λ>300 μm) the otherwise opaque terrestrial atmosphere prevents FIR observations from ground-based facilities (Fig. 1).

Since its first light in 2011, GREAT has collected science data on more than 175 flights. In its early configuration GREAT operated, in parallel, two single-pixel LHe/LN$_2$ cooled cryostats hosting heterodyne detectors in the 1.5 and 1.9 THz frequency bands. In the following years, thanks to the instrument's modular design, operation was extended to more and higher frequencies, and by 2015 there was a choice of dual-color observations among 4 frequency bands [4]. A major upgrade came in 2015 with the addition of the upGREAT Low Frequency Array [5] (LFA) and later in 2016 with the High Frequency Array (HFA) [6]. The LFA consists of two 7-pixel arrays, working in dual polarization mode, to ultimately cover the 1.9 to 2.5 THz atmospheric windows. The HFA is equipped with a 7-pixel array operating at 4.7 THz in single polarization.

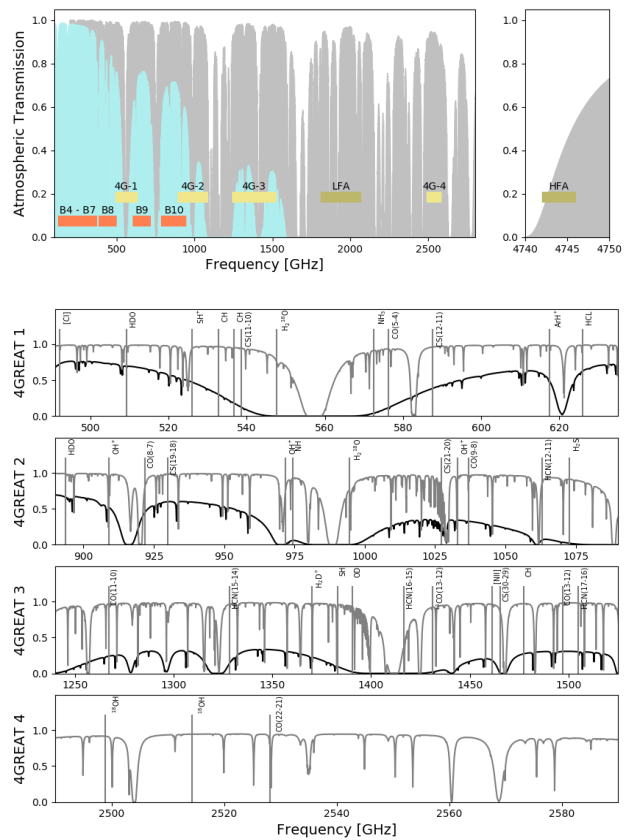

Fig. 1: The frequency coverage of the 4GREAT channels, of the upGREAT LFA and HFA arrays, and the ALMA receivers for reference, superimposed on the atmospheric transmission. For SOFIA the zenith transmission (grey) is calculated for a flight altitude of 13.1 km and 10μm of residual precipitable water; for the Chajnantor plateau at 5000m (ALMA, light blue) very best

Submitted on YYYYY xx 2020.

C. Durán, R. Güsten, B. Klein, O. Ricken, N. Reyes, H.-J. Wunsch and A. Görlitz are with the Max Planck Institut für Radioastronomie (MPIfR), 53121 Bonn, Germany (e- mail: cduran@mpifr-bonn.mpg.de).

J. Stutzki, U. U. Graf, K. Jacobs and C. E. Honingh are with the Cologne University, 50937 Köln, Germany.

C. Risacher is with the IRAM, 300 rue de la Piscine, 38406 Saint Martin d'Heres, France, and was with the MPIfR.

C. Durán is also with the European Southern Observatory (ESO), Alonso de Cordova 3107, Vitacura, Chile.

B. Klein is also with the University of Applied Sciences Bonn-Rhein-Sieg, 53757 Sankt Augustin, Germany (bklein@mpifr.de).

G. de Lange is with the Space Research Organization Netherlands (SRON Zernike Building, PO Box 800, NL-9700 AV Groningen, the Netherlands.

Y. Delorme and Jean-Michel Krieg are with the Sorbonne Université, Observatoire de Paris, Université PSL, CNRS, LERMA, F-75005, Paris, France.

D.C. Lis was with the Sorbonne Université, Observatoire de Paris, Université PSLtory/LERMA until 2019 and is now with the Jet Propulsion Laboratory, California Institute of Technology, 4800 Oak Grove Drive, Pasadena, CA 91109, USA.

---

[1] GREAT, the **G**erman **RE**ceiver for **A**stronomy at **T**erahertz frequencies, is a development by the MPI für Radioastronomie and the KOSMA/Universität zu Köln, in cooperation with the DLR Institut für Optische Sensorsysteme.

conditions of 200 µm PWV are assumed, though occurring for a few nights per year only. In the lower panels we zoom-in on the transmission for each of the 4GREAT bands (grey for SOFIA, black for Chajnantour), with a selection of important astrophysical transitions marked (compare with Table 1).

The addition of the arrays imposed major hardware changes. Because of the parallel operation of both arrays, the instrument was required to expand the IF processor and the digital back-end from 2 to 21 channels, each of them with frequency coverage from 0 to 4 GHz. To facilitate the array operation for astronomical observations, a K-mirror working as beam de-rotator, was installed as part of the instrument optics system. Finally, the use of closed-cycle coolers required the installation of adequate cooling infrastructure aboard the aircraft. The instrument status as of June 2017 is presented in [6].

With the success of the upGREAT arrays, and considering the enhanced infrastructure of the instrument and cooling capacity of the observatory, the idea of 4GREAT was born. The new instrument should facilitate the operation and scheduling by reducing the number of flight configurations, while further extending SOFIA's science capabilities. The requirements towards the design of 4GREAT were:
- to allow for the integration of the GREAT single-pixel units operating in the 1.2-1.5 THz and 2.49-2.52 THz frequency bands [4],
- and to extend the science opportunities to sub-THz frequencies, not covered by SOFIA after the cancellation of the CASIMIR project [7].

This paper describes the design of the instrument along with its integration, testing and commissioning on board of the SOFIA observatory. We will show that making use of the improved performance of today's dichroic (frequency) filters, up to 4 frequency bands can be arranged into a single cryostat for simultaneous science operation with co-aligned pixels on the sky. The final choice of frequency bands is presented in Table I. For the lower-frequency extensions we gave preference to those frequency regimes that cannot be observed from ground-based facilities, thereby taking unique advantage of the transparent atmosphere at SOFIA's flight altitude.

To exemplify the new science opportunities: absorption line spectroscopy towards FIR background continuum sources will allow in-depth studies of the galactic interstellar medium with hydrides (many of which have their ground-state transition in the lower 4GREAT bands, see Table 1) – this will be a unique addition to what has commenced so successfully with HIFI/Herschel [8], but now operating up to 5 frequency bands simultaneously.

The development was launched at the end of 2015, as part of the PhD thesis work of C. Duran [9][10], and 18 months later 4GREAT (with 3 of its 4 channels) was integrated with the GREAT infrastructure. Its first-light flight took place on July 13, 2017 out of Christchurch, New Zealand.

TABLE I
4GREAT FREQUENCY COVERAGE

| Channel | Frequency (GHz) | Astrophysical lines of interest (examples) |
| --- | --- | --- |
| 4G-1 | 491 - 635 | [CI], **HDO**, **CH**, $^{(13)}$**CH**, **H$_2^{18}$O(1$_{10}$-1$_{01}$)**, **NH$_3$**, CO, **ArH$^+$**, **HCl**, SH$^+$, CS$^a$ |
| 4G-2 | 890-1090 | $^{(13)}$CO, HDO, H$_2^{18}$O(2$_{02}$-1$_{11}$), H$_2$S, OH$^+$, **NH**, **NH$_2$**, CS$^a$ |
| 4G-3 | 1240 - 1525 | [NII], **OD**, $^{(13)}$CO, H$_2$D$^+$, **SH**, CH, CS$^a$ |
| 4G-4 | 2490 - 2590 | $^{(18)}$**OH**, $^{(16)}$**OH**, CO(22-21) |

Notes: The usable sky frequency coverage is calculated for an IF center frequency of 6 GHz (SIS) and 1.5 GHz (HEB), respectively. CO transitions are accessible in all bands, the superscript $^{(13)}$ indicates that at least one isotopic transition can be observed. Script $^{(a)}$ denotes that multiple transitions of linear rotors like CS, HCN, HNC, HCO$^+$ and SiO can be observed in bands 4G-1 to 4G-3. In bold we mark molecular ground-state transitions.

## II. THE 4GREAT DESIGN

The 4GREAT design follows the modularity of the GREAT instrument as described in detail in [2]. There are four clearly distinguishable hardware groups:
- The 4GREAT receiver cryostat
- The local oscillator "upper" unit (LO-U), hosting the 4G-1 and 4G-2 LO chains and guiding optics, located on top of the optic compartment.
- The local oscillator "lower" unit (LO-L), hosting the 4G-3 and 4G-4 LO chains and guiding optics, located below the optic compartment.
- The 4GREAT optics plate (common warm optics)

These four modules are installed in the main instrument frame, along with the common support and control hardware. The latter includes a calibration unit (cold and hot load, with optics), the mixer bias controller, LO synthesizers, readout electronics for temperature sensors, and control for motorized parts of the optics (grid attenuators, calibration unit, K-mirror). Fig. 2 shows the disposition of the different modules as mounted to the instrument structure. We discuss the individual modules in the following sections.

*A. The 4GREAT Detectors*

The 4GREAT cryostat (Sec. II.E) hosts the mixer for each of the four channels, the low noise amplifiers (LNA), the isolators, temperature sensors, and the cold part of the optics. All of them are mounted together in a unique structure, named *cold tower*, made of aluminum (Fig. 3). The detectors are of different technologies: The 4G-1 and 4G-2 channels utilize Superconductor-Insulator-Superconductor (SIS) mixers, whereas 4G-3 and 4G-4 employ Hot Electron Bolometers (HEB). All mixers are of Double Side Band (DSB) response. Table II summarizes the main characteristics for each mixer.

*1) SIS Mixers for 4G-1 and 4G-2*

The SIS mixers for 4G-1 and 4G-2 are units originally developed for HIFI, the heterodyne instrument [11] onboard the Herschel Space Observatory. They provide an output IF signal that runs from 4 to 8 GHz.

The 4G-1 mixer is the spare flight mixer developed by LERMA in Paris, France, and therefore complies with all the specifications for a HIFI band 1 unit. Mixer noise temperatures lower than 100 K DSB have been reported for operation at 2 K bath temperature [12], a temperature that cannot be provided by the cryocoolers aboard SOFIA. As this good mixer performance is dependent on the actual junction temperature, our mechanical design places the mixer as close as possible to the coldest point of the 4K stage. The horn is thermally connected to the cold finger by flexible copper straps, reaching an operation temperature of 3.45K.

Similarly, the 4G-2 mixer is a flight spare mixer of the HIFI band 4, developed by SRON [13]. In 4GREAT we extend the operation of the mixer well below its design frequency range to provide access to important astrophysical lines (Table II), while compromising the unit's noise performance.

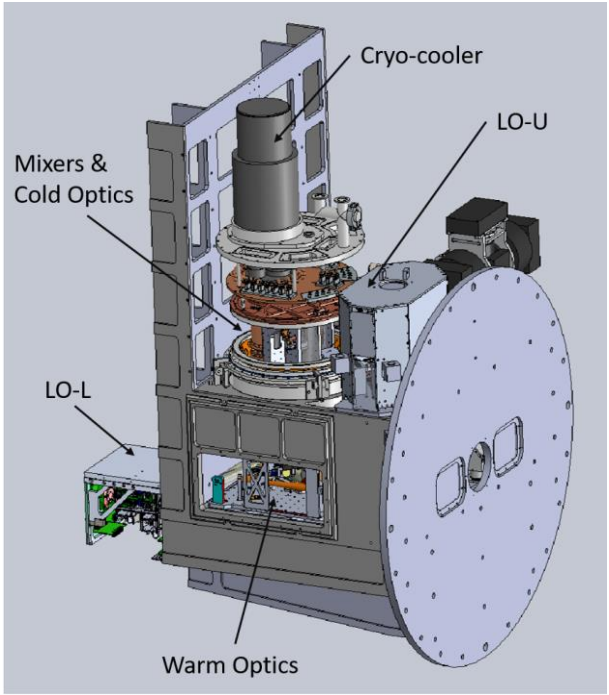

Fig. 2: Overview of the main 4GREAT modules mounted in the GREAT structure. Some panels and sidewalls, including cryostat vessel, have been removed for better visualization.

Both mixers have their bias network incorporated in the housing, as well as the coil to generate the magnetic field to suppress the Josephson effect, and a heater to temporarily increase the junction temperature and remove trapped magnetic flux if needed. The connection between each SIS mixer and its respective LNA is by a ferrite isolator, to minimize the effect of standing waves due to impedance mismatch.

TABLE II
4GREAT MIXER CHARACTERISTICS

| Band | Built by | Technology | RF/IF band (GHz) |
|---|---|---|---|
| 4G-1 | LERMA | SIS Nb [12] | 480-640 / 4-8 |
| 4G-2 | SRON | SIS NbTiN [13] | 960-1120 / 4-8 |
| 4G-3 | KOSMA | HEB NbTiN [14] | 1200-1600 / 0.1-2.6[1] |
| 4G-4 | KOSMA | HEB NbN [15][16] | 2500-2700 / 0.1-3.5[1] |

Note: [1] The HEB receiver temperature increases with IF, the 3dB roll-off bandwidth is quoted.

*2) HEB mixers for 4G-3 and 4G-4*

Both 4G-3 and 4G-4 detectors were originally developed by the KOSMA group at the Cologne University for the single pixel cryostats GREAT-L1 and GREAT-M, respectively. The 4G-3 mixer performance and technology are described in [14]. Per FTS response measurements the mixer performs between 1.2 and 1.6 THz (3dB fall-off). The mixer includes a corrugated horn clamped to the mixer body. A coaxial SMA connector is used for the IF output, and also used to supply the bias voltage through an external bias-T, which is integrated with the LNA. The 4G-4 HEB is fully described in [15][16]. It uses a novel custom-scaled spline-profile conical feed [17][18] which was micro-machined by Radiometer Physics GmbH (Meckenheim, Germany). The NbN HEB mixer has been optimized to cover the RF band from 2.5 to 2.7 THz. Like 4G-3, the 4G-4 mixer requires an additional biasing module included in the LNA.

*B. Local Oscillators*

4GREAT makes use of four fully independent, tunable local oscillator units [19], made up of cascaded multiplication stages and a high frequency power amplifier driver. The chains have been custom manufactured by Virginia Diodes Inc. (Charlottesville, USA). The reference input signal, provided by external oscillators, is between 10 and 16 GHz with a minimum power of 10 dBm. The four chains are grouped in pairs, in two separated enclosures, due to physical constraints. The 4G-1 and 4G-2 LO chains are lodged in the LO-U, whereas the 4G-3 and 4G-4 LO are in the LO-L. Table III summarizes the LO performance for all four chains.

*1) The LO-U*

The LO-U unit is placed above the optics compartment of the GREAT structure, in front of the 4GREAT cryostat. The wide RF bandwidth of the 4G-1 LO chain, ranging from 495 GHz to 630 GHz, is realized by operating two frequency-shifted driver stages. The resonance of the diplexer that combines the two signals has been tuned to match the strong atmospheric absorption by the ground-state water line (gap between 550 and 565 GHz). The chain delivers at least 200 µW and more than 300 µm across 95% of its bandwidth. In a similar hybrid approach the 4G-2 chain, also operating two driver stages, generates a signal output larger than 200 µW from 850 to 975 and 990 to 1085 GHz. Again, the gap matches strong telluric water absorption.

Apart from the LO chains for 4G-1 and 4G-2, the LO-U unit contains parts of the optics required to guide the LO beams towards the common optics plate inside the optics compartment (at aircraft external pressure, see below). Hence, when in flight, the LO-U is part of the aircraft pressure boundary, and the optical LO path has to allow for vacuum windows. These windows, made of silicon with an antireflection coating, are identical to the vacuum windows of the cryostat.

*2) The LO-L*

The LO-L unit is placed in the compartment assigned for the local oscillator source servicing the right-hand side (looking towards the telescope) GREAT cryostat, located underneath the instrument structure (Fig. 2). It contains the chains for the 4G-3 and 4G-4 channels and their respective fore-optics. Additionally, the compartment hosts the power supplies for all four LO systems.

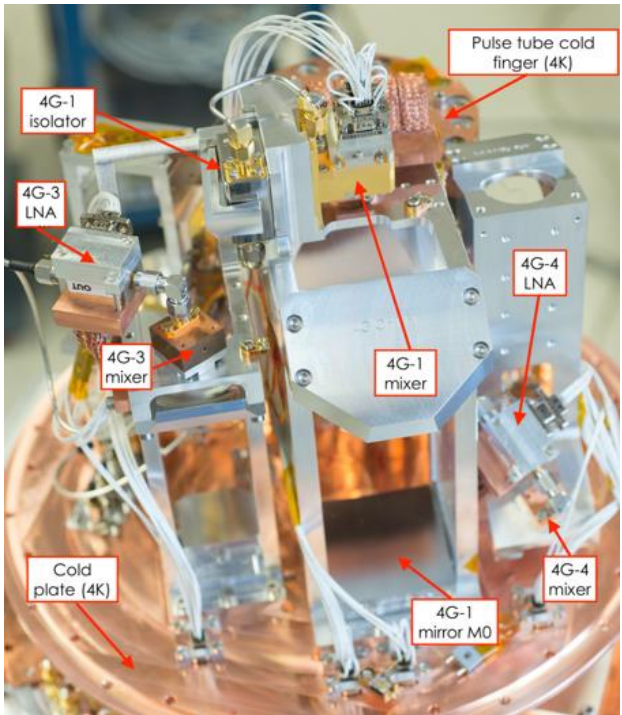

Fig. 3: The 4K stage of the 4GREAT cryostat. The cold towers mount the mixer, the LNAs, part of the optics and temperature sensors for each channel.

The 4G-3 LO follows again a hybrid approach to cover the wide RF band from 1240 to 1525 GHz. The gap is tuned to match the strong atmospheric absorption between 1395 and 1425 GHz. This LO chain delivers a minimum power of 40 µW over its operational range.

Unlike the other three LO sources, the 4G-4 LO is a single multiplication chain. With a multiplication factor of 216 the chain provides more than 3.5 µW of output between 2480 and 2620 GHz.

TABLE III
4GREAT LOCAL OSCILLATORS PERFORMANCE

| Channel | LO BANDWIDTH (GHz) | POWER[1] (µW) | MULT. FACTOR |
|---|---|---|---|
| 4G-1 | 495-550 / 565-630 | 300 | 48 |
| 4G-2 | 850-975 / 990-1085 | 200 | 72 |
| 4G-3 | 1240-1395 / 1425-1525 | 40 | 108 |
| 4G-4 | 2480-2620 | 3.5 | 216 |

Note: (1) minimum output power across >95% of the LO bandwidth, as measured by the manufacturer using a bolometric power meter.

*3) Reference synthesizer*

The GREAT reference signal box has been upgraded to contain four independent LO reference signals. They are made up of a digitally controllable YIG based synthesizer and a 30 MHz wide tunable YIG filter. The synthesizer and the bandpass filter can be tuned from 8 to 20 GHz. The YIG filter is used for harmonic filtering, eliminating unwanted harmonics and spurious signals from the synthesizer. It also allows performing a smooth ramp up of the LO reference signal, by slowly tuning the YIG filter to match the requested frequency.

*C. Optics Design*

Some of the relevant SOFIA telescope optical characteristics [20] are summarized in Table IV. The nominal telescope focal plane is placed 300 mm behind the Science Instrument interface flange (towards the cabin). The optical image de-rotator that has been installed with the upGREAT arrays, adds 319.7 mm to the signal path (the K-mirror is located in the optics path common to all channels). This change of the telescope's effective focal length is compensated by adjusting the position of the telescope's sub-reflector.

All 4GREAT channels are designed for an edge taper of 14 dB on the telescope secondary mirror. The optics setup uses only reflective elements except for the dichroic filters and polarization grids needed for simultaneous operation of the four channels and the co-mounted upGREAT high frequency array.

TABLE IV
SOFIA - OPTICAL PARAMETERS WITH GREAT

| | |
|---|---|
| Primary diameter | 2500 (2700) mm |
| Primary focal length | 3200 mm |
| Secondary diameter | 352 mm |
| Telescope total focal length | 49141 mm |
| Focal length with K-mirror | 51421 mm |

Note: the clear aperture of the oversized primary mirror is 2500 mm

*1) Signal Path Optics*

The first optical element in the signal path (Fig. 4) is a low pass dichroic filter from QMC (Cardiff, UK), used to separate the beams of the HFA and 4GREAT. The dichroic has a cut-off frequency, $\nu_c$, of 3.3 THz, its frequency dependent transmission varies between 88% and 98% over the 4G operation range. The signal loss in the HFA path is low, as the reflectivity is 97% at the frequency of the [OI] 63 µm fine-structure transition, i.e. the operating frequency of the HFA. The dichroic is located in the focal plane of the instrument, attached to the HFA optics plate.

The signal optics for each of the 4GREAT channels is made of the same kind of elements, but dimensioned for their respective frequency ranges. Some of these elements are common to two or even all of the channels. The 4G-4 optics has additional components due to its different LO coupling scheme (see below).

The next element that the signal encounters, after the first dichroic filter, is an ellipsoidal mirror (named M4), which is common to all 4 channels. This mirror focuses and deflects the beam to a first wire-grid, WG, which separates the signal beam into two polarizations. Each of the new beams, namely horizontal and vertical, are then re-focused by active mirrors $M3_{13}$ and $M3_{24}$, respectively, which, along with M4, form the first Gaussian beam telescope of each channel. Next, the horizontally and vertically polarized beams are split in frequency by dedicated low-pass dichroic filters $D_{13}$ ($\nu_c$ = 1.1 THz) and $D_{24}$, (1.6 THz), yielding now 4 independent beams: horizontal-low (4G-1), horizontal-high (4G-3), vertical-low (4G-2) and vertical-high (4G-4). Each individual beam is subsequently reflected to a group of flat/active (F2/M2) mirrors which provide four degrees of freedom for the optical alignment of each beam. The beams then pass their respective LO coupling grids (an interferometric diplexer in the case of 4G-4, placed before the flat/active group) and a polarization cleaning grid, aligned with the respective mixer polarization, before entering the corresponding cryostat vacuum window. Inside the cryostat, the beams pass through an infrared filter located on the 45 K shield before reaching the cold optics.

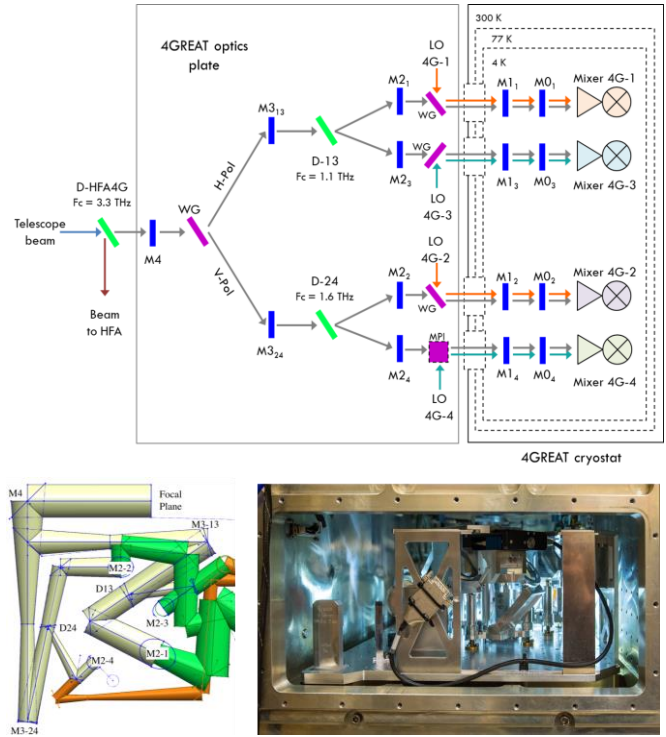

Fig. 4: Top. Sketch of the signal beam splitting by polarization and frequency-selective dichroics, and the active optical elements in each of the channels of 4GREAT. The upGREAT HFA and 4GREAT are operated simultaneously. In the bottom-left figure we illustrate the signal (golden) and LO (green and orange) beam propagation, in two levels. A side-view picture of the optics plate is shown at the bottom-right.

The cold optics is composed of a focussing mirror (M1), a flat mirror (F1) and a parabolic mirror (M0) in front of the respective detector horn. The pair of M2/M1 mirrors form the second Gaussian beam telescope in the beam path of every channel.

As an example, to illustrate the optic design, Fig. 5 displays the unfolded signal path for channel 4G-3, showing the critical components and the re-imaging with the Gaussian telescopes.

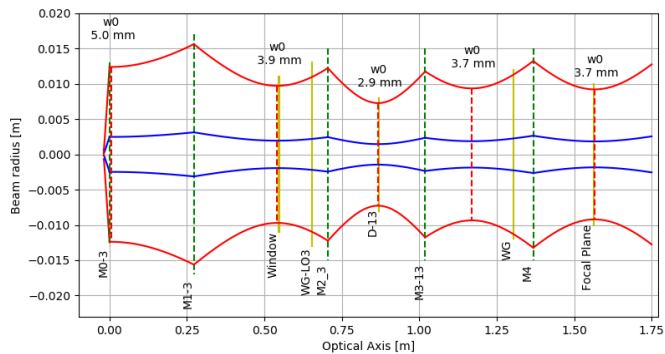

Fig. 5: Diagram of the unfolded optic signal path of the 4G-3 channel, at the design frequency of 1350GHz. $1w$ and $5w$ contours are shown in blue and red respectively. Beam waist positions are shown with dashed red lines. Yellow markers indicate the positions of beam-splitting optical elements and windows.

### 2) LO Coupling Optics

The 4G-1, 4G-2 and 4G-3 LO signals are optically combined with their respective sky signals by wire grids. The amount of LO and signal power that is coupled to the detectors depends on the polarization angle of the grid with respect to the polarization of the local oscillator and signal beams. Coupling more LO power has a penalty in the receiver temperature, as less signal from sky is coupled. The 4G-4 channel, due to its relatively low LO power, employs a Martin-Puplett interferometer as a diplexer.

As the power delivered for every chain is not uniform in frequency, an adjustable optical attenuator has been installed in the beam path of each LO. 4G-2, 4G-3 and 4G-4 use a concentric rotating polarizer (wire-grid). Therefore, the effective angle of the wires of the attenuator with respect to the LO coupling grid defines the amount of coupled power. 4G-1 uses a turntable variable aperture that truncates the LO beam, and with this its power.

### D. Front-end Cryostat and closed-cycle Cooler

The 4GREAT cryostat is an exact copy of those used for the upGREAT LFA and HFA. It has been fabricated by CryoVac GmbH (Troisdorf, Germany), and complies with all specifications on the GREAT vessels to certify airworthiness.

The receiver is cooled by a two-stage closed-cycle pulse tube refrigerator from TransMIT GmbH (Giessen, Germany). The pulse tube cooler provides about 0.8 W of cooling power at 4.0 K and 10 W at 45 K for the first stage. The pulse tube is connected to its rotary valve by a 0.75 m flexible hose. This helps to decouple the mechanical vibrations from the valve, located outside the cryo-vessel, from the pulse tube where the mixers are located. The frequency of the pulse tube is controlled by a three-phase variable frequency drive. The pulse tube stages are mechanically connected to the $1^{st}$ and $2^{nd}$ stage plates (and their respective shields) through flexible copper braids. The stages are held in position by fiber glass cylindrical structures.

In the outer vacuum walls of the 4GREAT cryostat are vacuum RF windows. The 4G-1 and 4G-2 windows are made of quartz with antireflection coating, supplied by QMC (Cardiff, UK). The 4G-3 and 4G-4 windows, manufactured by Tydex (St.Petersburg, Russia), are made of silicon with a parylene coating. Each of the 4GREAT channels uses one layer of Zitex G104 [21], cooled to about 45 K, as infrared filter. Table V summarizes the maximum and minimum window transmission, as well as the average across the band of the windows and filter for each band.

TABLE V
TRANSMISSION OF THE 4GREAT WINDOWS, FILTERS AND DICHROICS IN PERCENTAGE

| Channel | 4G-1 | 4G-2 | 4G-3 | 4G-4 |
|---|---|---|---|---|
| $1^{st}$ Dichroic | 93.9 | 96.7 | 94.8 | 88.4 |
| $2^{nd}$ Dichroic | 95.9 | 96.3 | 93.3 | 96.6 |
| Pol.splitter | 99.0 | 99.0 | 99.0 | 99.0 |
| LO coupler[1] | 97.0 | 91.0 | 93.0 | 93.0 |
| Window | 96.7 | 94.7 | 85.3 | 90.0 |
| IR-filter | 91.7 | 98.0 | 98.2 | 94.7 |

Notes. Numbers are averages across the respective RF tuning range, based on data provided by the manufacturer (QMC) or derived from our FTS measurements. [1] The loss depends on the available LO power at a given frequency.

The electrical wiring inside the cryostat is done by using phosphor-bronze wires. Capacitive filter networks are used to provide filtering of DC lines and better thermal isolation. Filters, integrated in D-sub and D-micro connectors are placed on the 300 K, 45 K, and 4K interfaces. The IF output signals are routed using flexible coaxial cables between the LNAs and the interface connectors on the 4K stage, and semi-rigid coaxial cables with stainless steel outer and copper-beryllium inner conductor between the 4K, 45K and 300K stages. The former allows easier servicing of the main components, the latter have been chosen because of their robustness, and to reduce thermal transfer between stages.

When the receiver is in operation, i.e. mixers and LNAs are powered on, and LO signal is applied to the mixers, the copper plate temperatures are 43 K for the first stage and 3.9K for the second stage.

*E. IF Signal Processing*

The IF output of each mixer is connected to an LNA. In the case of the SIS channels, 4G-1 and 4G-2, this is performed through ferrite isolators. These channels use cryogenic low noise amplifiers from Low Noise Factory (Gothenburg, Sweden), covering an IF band of 4-8 GHz with over 40 dB of gain. They have a noise temperature of only 2 K, when cooled below 6 K of physical temperature.

The 4G-3 and 4G-4 HEB output IF signals are amplified by SiGe cryogenic LNAs CITLF4 from Cosmic Microwave Technology, USA. The amplifier gain is 35-40 dB and the noise temperature is 4-5 K when measured at 23 K physical temperature. The LNAs were modified by replacing a 5 kΩ resistor used in an external bias network by custom-made coils. These amplifiers define the lower edge of the receivers IF range at 0.2 GHz,

The output of each LNA is routed, as explained above, to the cryostat top plate that accommodates the electrical feedthroughs: bias connectors, temperature sensors, heaters, and IF connectors. Pre-amplifier bias modules and warm IF amplifiers, with a gain of 30 dB, are enclosed in a case directly mounted on the top plate.

Each of the four IF outputs of the cryostat are connected to an individual module of the IF processor, located in the telescope's counterweight rack, by a 3-meter long flexible coaxial cable. The function of the IF processor is to filter, equalize, and adjust the power to the optimum input level required by the spectrometer. While the channels utilizing HEB detectors produce an IF output in the 0.2 to 4 GHz range, the 4G-1 and 4G-2 SIS channels operate from 4 to 8 GHz. To accommodate any required configuration between the HFA, the LFA, and 4GREAT, the IF-processor, built at the MPIfR, is therefore composed of 23 modules (21 modules: 0.1-4 GHz; 2 modules: 4-8 GHz) plus their controller. They are internally equipped with a total power detector used by the power auto levelling feature and to characterize the mixer conversion curves during instrument tuning.

Finally, the signals are fed into the MPIfR-built Fast Fourier Transform Spectrometers (FFTS, [22][5]) which digitize the signal and calculate the spectral power distribution of the detected signals. The spectrometer offers 11 dual-input independent boards, serving 0.1-4 GHz (of up to 22 HEB output channels), and three single input boards for the 4-8 GHz SIS bands. The latter achieve this IF coverage by fast input digitizers, which allow using the 2nd Nyquist band in the Fourier transform. Due to this feature, all of the IF processor channels could be designed without a frequency conversion stage. Although programmable to provide up to 64k channels, all of the spectrometer boards are currently configured for operation with 16k spectral channels only to limit the data rate for fast dumping observing modes. A spectral resolution of 244 kHz, corresponding to a velocity resolution of around 0.15 km/s at 492 GHz, the lowest 4G-1 observing frequency, and 0.015 km/s at the 4700 GHz [OI] frequency (HFA), meets all of our science requirements.

### III. 4GREAT INTEGRATION AND LABORATORY RESULTS

4GREAT was integrated and tested using the 4G-1, 4G-2 and 4G-4 detectors in the MPIfR laboratories between late January 2017 and late March 2017. In order to maintain the integrity of the GREAT receiver, still in-flight operation during those months, we decoupled the commissioning of these three units from the later addition of the 4G-3 (former GREAT-L1). 4GREAT was then shipped to the NASA Armstrong Flight Research Center (AFRC) in Palmdale, USA, for integration and alignment with the actual GREAT flight hardware. In the following, we describe critical steps in this integration process and present results achieved during this laboratory verification.

*A. Alignment of the GREAT optics*

*1) LO coupling and alignment*

The LO optics is designed to provide 4 degrees of freedom: tip, tilt and shift on the vertical and horizontal directions. The alignment and coupling of the local oscillator signal to each channel is optimized by manipulating the adjustable mirrors in this path, while inspecting the I-V curve of the respective mixer. During this process the angle of the LO coupling grid (hence, the LO power at the mixers) is also determined.

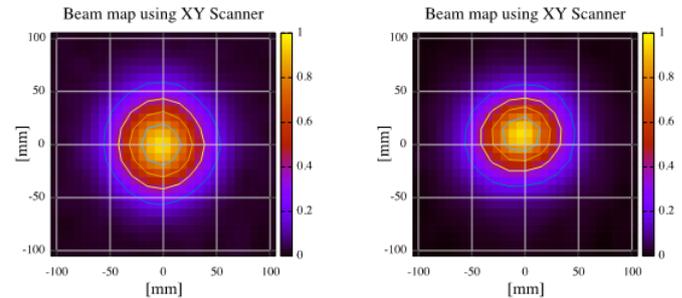

Fig. 6: Beam measurements for 4G-1 (left) and 4G-2 (right) using an external beam scanner. The measured 10mm offset translates in a small offset of 25mm at telescope secondary (diameter 350mm), causing a negligible difference in coupling to telescope. The data were taken during the channel alignment process of the GREAT flight configuration at the AFRC, USA.

*2) Signal beam alignment*

Signal beam alignment is a slow, iterative process. The basic principle is to measure the beam response at two distances along the signal path: from these, we calculate the beam positions in the focal plane (using the standard ABCD matrix method), and derive and apply the necessary changes to the positions of the adjustable warm optics elements. This sequence is performed iteratively until a satisfactory alignment of the optics is achieved. The relative beam positions on the sky can be derived from the measured offsets in the focal plane with the known telescope plate scale of 4″/mm, and are later-on verified on the sky.

The first signal beam response is measured close to the focal plane using a compact beam measurement wheel [23] installed within the optics compartment of the GREAT structure. A second beam measurement is done in the laboratory with an external 2-axes scanner, placed at 3000 mm from the SI flange, slightly less than half the distance to the secondary mirror. The measurement target on the scanner consists of a sheet of mm-wave absorber material at room temperature with a central hole. Scatter cones with varying free apertures can be mounted inside the hole. Behind the hole a flat mirror points towards absorber material floating in a dewar filled with liquid nitrogen.

The beam positions are determined by sweeping vertically and horizontally the target, meanwhile recording the total power counts from each channel. Once this alignment is done, a high-resolution map is taken to check for beam distortions, like asymmetric beams or coma and sidelobes. Fig. 6 shows the beam maps for two channels of 4GREAT during the alignment phase in the laboratory. Important parameters to evaluate are the instrument coupling to the sub-reflector, and the channels co-alignment. The latter is fundamental for making use of the simultaneous scientific capabilities, among the 4GREAT channels and the HFA array.

The final alignment verification in tilt can only be performed with the instrument mounted to the telescope flange by directly moving a test probe across the telescope's sub-reflector. In this procedure, portions of the secondary mirror of the telescope at predefined positions are covered with a small rectangular (8 x 25 cm) paddle of Eccosorb dipped into liquid nitrogen. Variations in the total power level for the selected channel are then correlated to the positions of the paddle, and the illumination of the subreflector can be derived. The final verification of the offsets is done by astronomical measurement on the sky (see below).

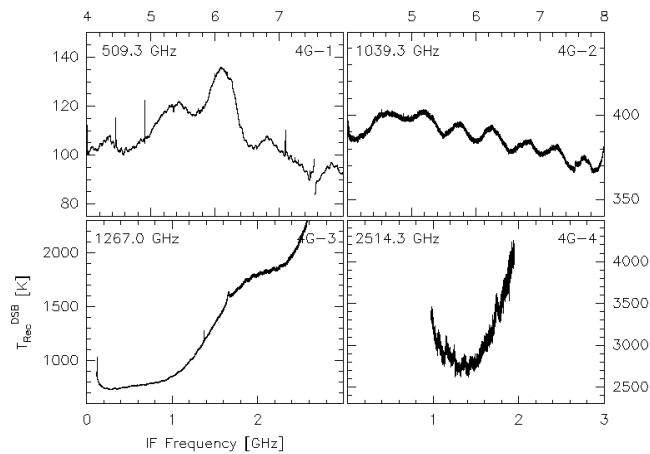

Fig. 7: Uncorrected DSB receiver temperature $T_{Rec}$ (K) versus IF frequency for all 4GREAT receivers as measured in flight during the November 2019 flight series. The sky frequency of the observation is given. For the 4GREAT channels operating SIS mixers we show the 4-8 GHz IF range, while for the HEBs the displayed IF is 0 to 3 GHz.

### B. Receiver characterization[1]
#### 1) Receiver Noise Temperatures

Receiver temperatures ($T_{Rec}$) of the complete instrument were measured for each channel at different LO frequencies, using the Y-factor method. The optics compartment was evacuated to avoid atmospheric absorption in the measurement, representative for conditions at flight altitude.

The cold load is implemented as an absorber cooled to around 70K by a Stirling cooler. An external $LN_2$ load was used to cross-reference the temperature scale of this internal calibration unit. Spectroscopic measurements of $T_{Rec}$ across the receiver IF were obtained with the FFTS spectrometer. Fig.6 displays representative results for all 4 channels. Values presented here are on the DSB temperature scale, as measured in the actual 4GREAT flight configuration.

TABLE VI
4GREAT RECEIVER PERFORMANCE

|  | 4G-1 | 4G-2 | 4G-3 | 4G-4 |
|---|---|---|---|---|
| RF Bandwidth[1] (GHz) | 491-635 | 890-1090 | 1240-1525 | 2490-2590 |
| IF Bandwidth (GHz) | 4-8 | 4-8 | 0.5-2.5 | 0.75-2.25[2] |
| $T_{Rec}$ (DSB) [K] | 100-140 | 300-600 | 1000-1200 | 3000 |

Notes: [1] The effective bandwidth includes a central gap due to LO features. Also consider atmospheric transmission features. [2] The 4G-4 IF bandwidth is limited by the MP diplexer to about 1.5 GHz, centered around any frequency in the range 1.4-2.9 GHz.

Results show that the receivers perform as expected on the basis of their RF design bands and LO coverage. 4G-1 can be operated with $T_{Rec}$ of 100-140 K (DSB) over much of the IF and RF bands. The small noise bump at mid-band is likely due to a mismatch between mixer and cold amplifier, and will be addressed in a future maintenance campaign. As is, science is little affected as the large IF bandwidth does allow shifting the line-of-interest to the sweet spots. At the low RF band edge, we verified on-sky operation of the CI 492 GHz transition with ~150 K (LSB tuned). 4G-2 performs well within its nominal design band (Table II), with $T_{Rec}$ ~300-400 K above 1 THz, rising to ~600 K at 950 GHz and below. 4G-3 performs with similar figures as during its operation in the "old" GREAT channel L1, with $T_{Rec}$ between 1000 and 1200 K across its operational range (typically measured at an IF frequency of 1.2 GHz). The noise performance of 4G-4 depends heavily on the tuning of the Martin-Puplett interferometer (in a trade-off between passband width and IF frequency, defined by the science requirements), but DSB temperatures of 3000 K can be achieved. We summarize the performance figures of 4GREAT in Table VI.

At this point, the latest, the interested reader may wonder about the competitive performance of the 4GREAT - with its color-splitting, complicated fore-optics - against a straight-forward direct approach with reflective optics only. As even the first-light version of GREAT split the incoming signal by polarization [2], the answer is given in Table VI: the signal gain of a stand-alone 4GREAT channel, without parallel operation of the HFA, is by the combined losses of the two dichroics (first and second row in Table V); the gain of a 4GREAT channel - still operating in parallel with the HFA - is reduced by the loss of the 2nd dichroic. To summarize, the 4GREAT fore-optics with its state-of-the-art dichroics offers simultaneous observations in up to 5 frequency bands, at the penalty of ~10-15% signal loss, but with a much simplified logistics for the SOFIA observatory.

#### 2) Receiver stability

The Allan-variance (AV) [24] characterizes the stability of a receiver system and ultimately the maximum phase integration time when observing a source on sky. The AV for 4GREAT was measured looking at the calibration cold load and with an equivalent noise bandwidth of 1.13 MHz. We derive spectroscopy AV times above 100 sec for all bands. The data was acquired under laboratory conditions, with stable room temperature, pre-warmed electronic (IF processor and FFTS modules), and no other radiation sources active other than those required for operation of the receiver. In flight, with changing environmental conditions slightly worse stabilities are to be expected. For science operation we limit the 4GREAT observing modes to phase times of less than 50 seconds.

---

[1] As the instrument is regularly upgraded (inter alia the local oscillator bandwidth coverage and output power are improved, as technology allows) the performance described in this section reflects the receiver status as flown during SOFIA observing cycle 7-I (Nov 2019).

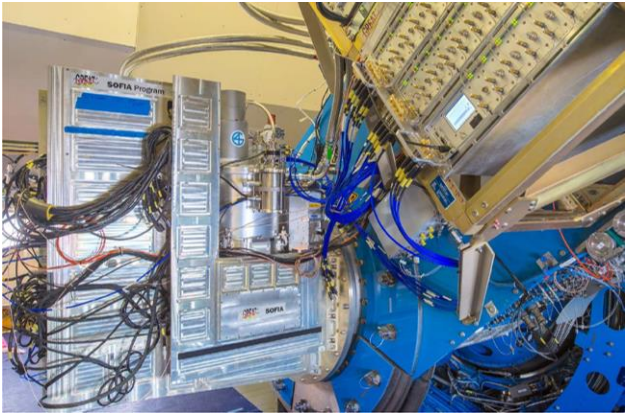

Fig. 8: The GREAT instrument in its HFA/4GREAT configuration, mounted to the SOFIA telescope flange. The 4GREAT cryostat can be partly seen on the right-hand side of the central instrument structure. The counterweight rack, with the back-end electronics is seen on the upper right, the rack on the left side houses the receiver control electronics.

## IV. COMMISSIONING AND FIRST LIGHT

The instrument's scientific commissioning took place in July 2017 during the SOFIA's 4$^{th}$ southern deployment to Christchurch, New Zealand. In June 2018 the 1.3 THz HEB mixer was transferred from GREAT-L1, bringing 4GREAT to its full capabilities. Fig. 8 shows a picture of the instrument installed on board of SOFIA.

4GREAT performed well during these verification flights, as did the HFA operating in parallel. Sensitivities were as measured in the lab prior to installation.

We determined the boresight and beam co-alignments by scans across and maps of Jupiter (2017) and Mars (2018), the only astronomical targets suitable for our optical verifications. Unfortunately, both planets were rather extended, compared at least to the higher frequency channels of 4GREAT. On July 11$^{th}$ 2017 Jupiter was extended with a 33.9 × 36.3″ disk diameter, while on June 28$^{th}$ 2018 Mars was extended with a 19.87″ disk diameter (in June 2017 Mars was a daytime target and therefore not accessible). Because the Martian atmosphere (and hence its spectral intensity distribution) is rather clean and well modeled (compared to Jupiter with strong atmospheric features imprinted to its continuum), the following numbers are based on the 2018 observations of Mars.

*1)* The HFA central pixel and the four 4GREAT channels are co-aligned within a circle of less than 1 arcsec radius. In flight, we track on the smallest beam requested by the science case, which is 4G-4 or (in most cases) the central pixel of the HFA. Boresight offsets of all pixels relative to the reference pixel are small compared to the beams (Table VII).

*2)* The observed beam patterns confirm the diffraction limited optics. The half-power full beam widths ($\Theta_{HPFW}$) presented in Table VII have been derived from cross-scans towards Mars, after deconvolving the observed profile from the emission of the Martian disk. An exception is 4G-4, for which the beam is too narrow to perform a reliable deconvolution, within the signal-to-noise of the data. We quote the width as extrapolated from the lower frequencies (10.5″).

*3)* Finally, comparing the observed vs. the modelled brightness of Mars [25] we determine the main beam coupling efficiency $\eta_{mb}$. The relatively low values are consistent with the expected values, due to the large central blockage by the tertiary folding mirror in SOFIA's folded Nasmyth optics.

In the course of these measurements we verified all the GREAT observing modes, from pointed to raster and on-the-fly observations, chopped and unchopped. 4GREAT performed well, meeting specifications. In Fig. 9 we present a typical result achieved during regular science observations of an astronomic target.

TABLE VII
GREAT BEAM PARAMETERS.

| Channel | #pixels | $\nu_{obs}$ (GHz) | $\Theta_{HPFW}$ (arcsec) | $\eta_{mb}$ |
|---|---|---|---|---|
| 4G-1 | 1 | 530 | 52 | 0.61 |
| 4G-2 | 1 | 1038 | 27 | 0.55 |
| 4G-3 | 1 | 1337 | 20 | 0.62 |
| 4G-4 | 1 | 2675 | 10.5 | 0.57 |
| HFA | 7 | 4745 | 6.3 (13.6) | 0.64 |
| LFA-H | 7 | 1900 | 14.1 (31.8) | 0.65 |
| LFA-V | 7 | 1900 | | 0.66 |

The half-power full beam width ($\Theta_{HPFW}$) and main beam coupling efficiency $\eta_{mb}$ was derived from observations of Mars at frequency $\nu_{obs}$. Figures for the LFA and HFA are from [6], numbers in parenthesis describe the pixel spacing in the hexagonal arrays

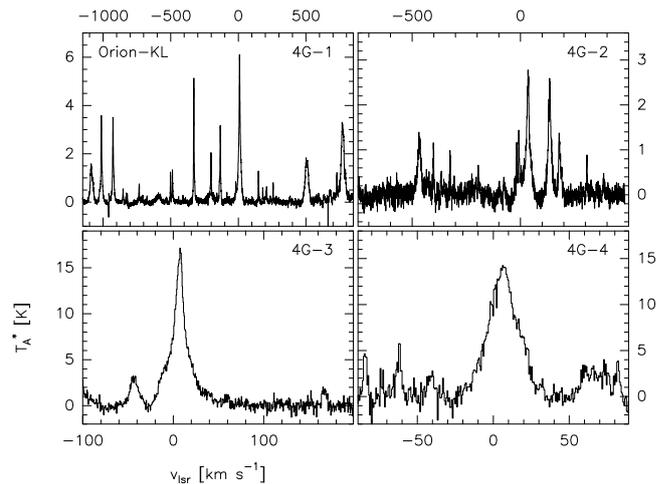

Fig. 9: Composite of 4GREAT spectra towards the well-known Orion-KL nebula, the nearest example of a high-mass star forming core. All spectra were observed simultaneously, with an on-source integration time of only 100 sec. The science case for this project asked for the $NH_3$ ground-state transition (572.5 GHz) to be observed with 4G-1, $NH_2$ (952.5 GHz) in 4G-2, the J=13-12 rotational transition of the $^{13}CO$ isotopologue (1431.2 GHz) in 4G-3, and the high-excitation $^{12}CO$(22-21) rotational transition in 4G-4 (2528.2 GHz). These lines are centered at the source velocity of 6 km/s. The unique chemical complexity of Orion-KL is reflected in the many other lines detected in the reception bandwidth of 4GREAT. Spectra are presented on the antenna temperature scale $T_A^*$, not considering source coupling efficiencies.

## V. Conclusion

We present details of the design, construction, and laboratory tests of the 4GREAT receiver. After its successful integration and commissioning in July 2018 the instrument has been made available to the SOFIA user communities with the call for proposals for observing cycle 6.

With the addition of 4GREAT, GREAT is currently composed of 3 cryostats: the LFA with its dual-polarization 7+7 pixels, the HFA with its 7 pixels single-polarization array, and 4GREAT with four independent single-pixel channels. The three cryostats are grouped in two flight combinations: The first combination, commissioned in June 2017, operates the LFA and HFA simultaneously. In the second combination, the four channels of 4GREAT observe in parallel with the HFA. Fig. 10 illustrates the beam sizes and pixel spacing of this configuration, superposed on an image of the Horsehead nebula.

4GREAT adds new scientific opportunities to the suite of SOFIA instrumentation. Its frequency-multiplexing will allow for a more efficient use of the precious observing time on board SOFIA. 4GREAT's extension to lower frequencies will re-open access to, among others, the ground-state transitions of light hydrides (Table I) studied so successfully with Herschel/HIFI [11].

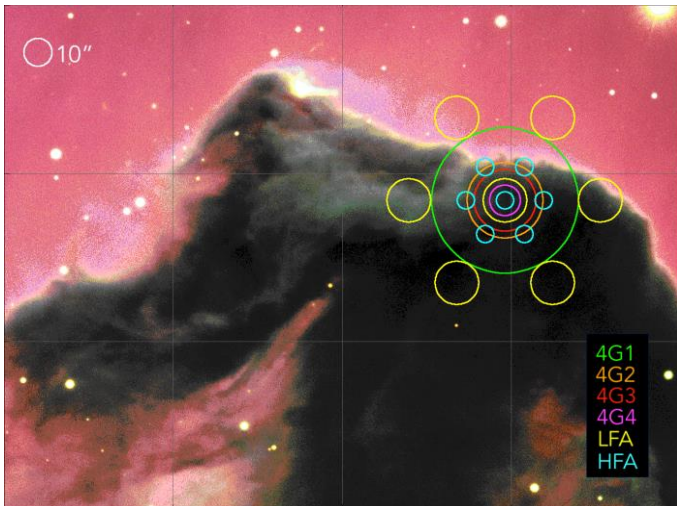

Fig. 10: The diffraction limited beams of the GREAT receivers 4GREAT, HFA and LFA, superposed on a 3x4 arcmin sized image of the Horsehead nebula, taken with the FORS2 instrument on the Very Large Telescope (ESO press release, www.eso.org)

### Acknowledgment

The authors thank the mechanical workshop at MPIfR, Bonn, for the excellent fabrication quality. The success of 4GREAT would not have been possible without the strong commitment of Virginia Diodes Inc. in the development of customized LO sources. Similarly, we thank QMC for the development of dichroic filters, and TransMIT for the pulse tube development and outstanding support before and after the instrument installation.

We thank the SOFIA engineering and operations teams for their strong continuous support during the 4GREAT installation, commissioning and later operation.

SOFIA is jointly operated by the Universities Space Research Association, Inc. (USRA), under NASA contract NNA17BF53C, and the Deutsches SOFIA Institut (DSI) under DLR contract 50 OK 0901 to the University of Stuttgart.

The development and operation of 4GREAT was financed by resources from the MPI für Radioastronomie, Bonn and the Universität zu Köln, and by the Deutsche Forschungsgemeinschaft (DFG) within the grant for the Collaborative Research Center 956 as well as by the Federal Ministry of Economics and Energy (BMWI) via the German Space Agency (DLR) under Grants 50 OK 1102, 50 OK 1103 and 50 OK 1104.

Mixers for Channel 1 of the 4GREAT (4G-1) instrument have been designed and developed by LERMA (Observatoire de Paris, CNRS, Sorbonne Université, Université de Cergy-Pontoise) in the framework of the Herschel/HIFI project, with funding from CNES.

Part of this research was carried out at the Jet Propulsion Laboratory, California Institute of Technology, under a contract with the National Aeronautics and Space Administration.

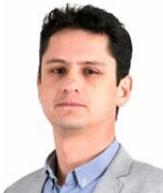

**Carlos A. Durán** received the Engineering and the M.Sc. degree from Pontificia Universidad Católica de Chile in 2004, and the Ph.D. degree in Astrophysics from Rheinischen Friedrich-Willems-Universität Bonn, in Germany, in 2018. Between 2004 and 2015 he worked as electronics engineer for the Atacama Pathfinder Experiment (APEX). From 2015 until 2018 he developed the 4GREAT instrument for the SOFIA airborne observatory as the main part of his PhD Work. After that, worked as Physicists for the MPIfR Submm technology division developing and servicing existing instrumentation for SOFIA and APEX. Since February 2020 he became the APEX Station Manager.

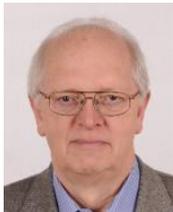

**Rolf Güsten** received his PhD from the University of Bonn in 1983. Until his retirement from active management duties end of 2018 he served as head of the Division for Submm Technologies at the MPI für Radioastronomie, Bonn (1996-), was the Principal Investigator for the development of GREAT, the heterodyne spectrometer operated on board of SOFIA (1998-) and the Project Manager responsible for the commissioning and early operation of the APEX telescope in Chile. Nowadays he is working as senior scientist at the MPIfR, reviving his interests in the field of submillimeter/far-infrared astrophysics to which he has contributed more than 250 papers in refereed journals, so far.

**Christophe Risacher** received the Engineering M.Sc. degree from l'École Supérieure d'Électricité (Supélec), Gif-sur-Yvette, France in 1998, and the Ph.D. degree from Chalmers University of Technology, Göteborg, Sweden, in 2005. Since 1998, he had worked at various radio astronomy observatories developing novel instrumentation and supporting observations. Among those are the IRAM 30m telescope in Granada, Spain, the Chalmers University of Technology with the Onsala Observatory, Sweden, the Apex Telescope with the European Southern Observatory, the HIFI instrument with the Herschel Observatory. From 2011 to 2018, he was working at the Max Planck Institut für Radioastronomie in Bonn, Germany as the project manager responsible for the development of the upGREAT array receivers for the SOFIA NASA/DLR airborne observatory. He is since 2018 the receiver group leader at IRAM in Grenoble, France.

**Bernd Klein** received the B.Eng. degree in electronics from the University of Applied Science Friedberg, Germany, in 1993, and the M.Eng. degree in theoretical electrical engineering from the University Siegen, Germany, in 1996, and the Ph.D.degree in astronomy from the University Bonn, Germany, in 2005. From 1999, he was an Engineer and, from 2002 to 2009 the head of the laboratory for digital technology at the Max Planck Institute for Radio Astronomy (MPIfR) in Bonn, Germany. In 2009, he became a Professor of digital signal processing and radio astronomical instrumentation at the University of Applied Sciences Bonn-Rhein-Sieg (H-BRS). Since September 2018 he has additionally taken over the management of the submm technology division at the MPIfR and since October 2018 he is also CoPI of GREAT at SOFIA.

**Hans Joachim Wunsch** photograph and biography not available at time of publication

**Andrej Görlitz** photograph and biography not available at time of publication

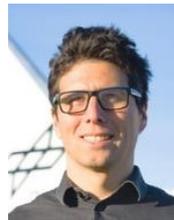

**Nicolas Reyes** received the Ph.D. degree in Electrical Engineering from Universidad de Chile in 2012. From 2008 to 2012 he worked in the design of the ALMA Band 1 instrument (35-50GHz). In 2013 he joined the Max Planck Institute for Radio Astronomy, where he worked in the development of UpGREAT for SOFIA observatory. In 2015, he joined Universidad de Chile as Assistant professor at the Electrical Engineering department. His research interests include microwave circuits, low-noise electronic, antennas, numerical simulation, optic and radio astronomy instrumentation. Since 2019 he is with the Max Planck Institute for Radio Astronomy as instrument scientist in the sub-milimetric technology division.

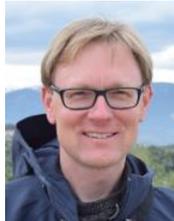

**Oliver Ricken** received his M.Sc. degree in Physics from the Rheinisch-Westfälische Technische Hochschule, Aachen, Germany in 2007, and his PhD degree in Experimental Physics from University of Cologne in 2012. During his PhD he started working with the GREAT instrument, a heterodyne spectrometer for SOFIA, and built up its 1.4 THz channel (GREAT-L1). Since 2012, he is an Instrument Scientist in the Sub-mm Technologies Division at the Max-Planck-Institut für Radioastronomie, Bonn, Germany, where he is developing new instrumentation and software tools for GREAT, as well as supporting observations with the instrument.

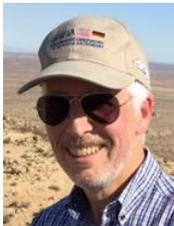

**Karl Jacobs** Karl Jacobs received the Diploma Degree in physics and astronomy from the University of Bonn, Germany in 1981 and the Dr. rer. nat. degree from the University of Cologne in 1984. In 1988 he founded the Superconducting Mixers group including a microdevices fabrication lab at KOSMA, University of Cologne. The group has developed and fabricated superconducting heterodyne mixers throughout the mm- submm and Terahertz range for instruments such as the KOSMA observatory, the AST/RO telescope at Amundsen-Scott South Pole Station and the Herschel Space Observatory. The lab provided all Terahertz mixers for the GREAT instrument on SOFIA, including the 4.7 THz array.

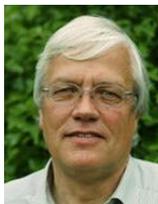

**Jürgen Stutzki** received his PhD from the University of Cologne in 1984. After a 2-year postdoc period at UC Berkeley, he became a senior research assistant at the MPI für Extraterrestrische Physik, Garching. In 1990 he became Assistant Professor for Physics at the University of Cologne and was promoted to full Professor in 2001. He has led the Submm-Astronomy group in Cologne since then, participating as co-Principle Investigator in the HIFI instrument on Herschel and developing the GREAT heterodyne instrument for SOFIA as co-Principle Investigator. With the retirement of the former PI, Rolf Güsten, he now serves as the PI of GREAT. His astrophysics research interest is in the physics and chemistry of the Interstellar Medium and Star Formation, which also drives his interest in developing novel astronomical instrumentation. He has contributed more than 200 papers in refereed journals.

**Urs U. Graf** photograph and biography not available at time of publication

**Cornelia E. Honingh** photograph and biography not available at time of publication biography.

**Gert de Lange.** photograph and biography not available at time of publication biography.

**Jean-Michel Krieg** received the Engineering M.Sc. degree from ENSMM Besançon, France in 1983, and the Ph.D. degree from University of Besançon, France in 1987. From 1987 to 1996 he worked at Thales Electronic Devices as a research engineer. He then joined Observatoire de Paris where he was involved in several space projects: ROSETTA-MIRO, HERSCHEL-HIFI and JUICE-SWI.

**Yan Delorme** has received the PhD degree in Microwave Electronics from Université Paris-Sud in 1993. Between 1993 and 1996, she worked at the French National Institute of Nuclear and Particle Physics on research and development of particle detection systems. Since 1997, she has been a research engineer at LERMA, Paris Observatory. Her main fields of interest are heterodyne detection techniques and applications in the submillimeter and THz range.

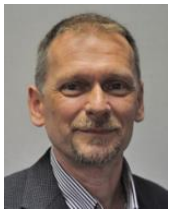 **Dariusz C. Lis** received his Ph.D. from the University of Massachusetts, Amherst in 1989. He was Senior Research Associate in Physics at the California Institute of Technology and Deputy Director of the Caltech Submillimeter Observatory, Professor and Director of Laboratory for Studies of Radiation and Matter in Astrophysics and Atmospheres at the Sorbonne University/Paris Observatory, and is currently Scientist at the Jet Propulsion Laboratory, California Institute of Technology. He is an expert in high-resolution molecular spectroscopy of the Solar System objects and the interstellar medium, from the Milky Way to the high-redshift Universe. Using ground-based and space-borne submillimeter facilities, including the Caltech Submillimeter Observatory, *Herschel* Space Observatory, and Stratospheric Observatory for Infrared Astronomy, he has studied molecules and their isotopic ratios as tracers of the physical conditions and chemistry in the interstellar medium, and the origin of Earth's water.